\documentclass[prb]{revtex4}
\usepackage{graphicx}

\begin{document}

\title{Nonlocal entanglement in hybrid superconducting and normal-metal \\three terminal devices}

\author{Jian Wei and V. Chandrasekhar}

\affiliation{Department of Physics and Astronomy, Northwestern University, Evanston, IL 60208, USA}

\date{\today}

\maketitle 

\parskip 7.2pt

\textbf{Nonlocal entanglement is crucial for quantum information processes. While nonlocal entanglement has been realized for photons, 
it is much more difficult to demonstrate for electrons. One approach that has been proposed is to use  hybrid superconducting/normal-metal devices~\cite{Lesovik2001,Recher2003}, where a Cooper pair splits into spin-entangled electrons in two spatially separated normal-metal leads. This process of nonlocal Andreev reflection is predicted to lead to a negative nonlocal resistance and positive current-current correlation~\cite{Byers1995,Deutscher2000}. By  cross-correlation measurements as well as measurements of the local and nonlocal resistance, we present here experimental evidence showing that by independently controlling the energy of electrons at the superconductor/normal-metal interfaces, nonlocal Andreev reflection, the signature of spin-entanglement, can be maximized.}

Multiple particle  entanglement has been realized for photons~\cite{Zhao2004} and has been used for quantum cryptography. However, for electrons, which are massive fermions, nonlocal entanglement has not been clearly demonstrated. Entanglement for electrons may arise in the spatial degree of freedom (orbital entanglement) or the spin degree of freedom (spin entanglement). Recently, orbital entanglement in a fermionic Hanbury Brown and Twiss two-particle interferometer was observed using current cross-correlation measurements~\cite{Samuelsson2004prl,Neder2007nature,Neder2007prl}, but further investigation is still required to verify whether the measured correlation is truely due to entangled states~\cite{Samuelsson2009}. Spin entanglement has been predicted to exist in hybrid superconducting/normal-metal (SN) devices~\cite{Lesovik2001,Recher2003}.  In this case, a Cooper pair splits into two spin-entangled electrons in spatially separated normal-metal leads in a process called nonlocal Andreev Reflection (AR)~\cite{Byers1995,Deutscher2000}, which leads to a positive current-current correlation and negative nonlocal resistance. Although a cross-correlation measurement in these systems would provide direct evidence of entanglement~\cite{Bignon2004}, previous experimental attempts focused on nonlocal resistance measurements~\cite{Beckmann2004,Russo2005,Cadden-Zimansky2006,Kleine2009}.  Here we show evidence of nonlocal entanglement from cross-correlation measurements as well as measurements of the local and nonlocal resistance. By independently controlling the energy of electrons at the superconductor/normal-metal interfaces, we show that nonlocal Andreev reflection, the signature of spin-entanglement, can be maximized.

Spin-entanglement at the SN interface can be understood in the context of AR~\cite{Andreev1964}, in which a low energy electron in the normal metal  impinges on the SN interface and a hole is retroreflected while a Cooper pair is created in the superconductor. When two normal metals are coupled to a superconductor with spatial separation comparable to the superconducting coherence length ($\xi_{S}$), roughly the size of a Cooper pair, it is predicted that electrons in the two normal metals can also be entangled via a nonlocal analog of AR called crossed Andreev reflection (CAR)~\cite{Byers1995,Deutscher2000}. As a Cooper pair splits into two entangled electrons that are then injected into the two normal-metal leads, instantaneous currents of the same sign are generated across the two SN interfaces, giving rise to a negative nonlocal resistance as well as a positive current-current correlation between the SN junctions. 

\begin{figure}
\includegraphics[width=5in]{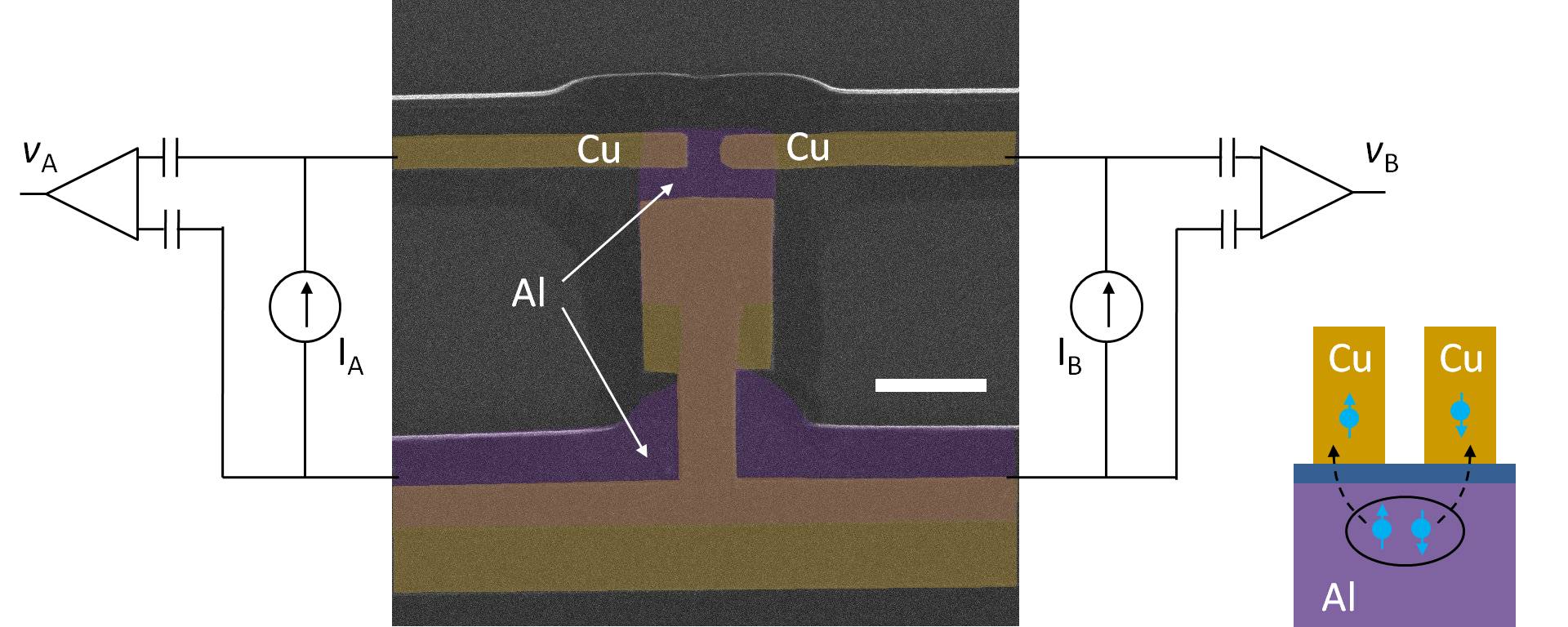}
\caption{ Schematic of the cross-correlation measurement. Scanning electron micrograph with false color enhancement of a three-terminal superconductor/normal-metal device. The scale bar is 1 $\mu$m. The area colored in dark purple is Al film, and the area colored in gold is Cu film. The thin white stripes are due to Al film deposited on the sidewall of the lithography mask. The two Al/AlO$_{x}$/Cu junctions are each biased with a dc current source and the ac voltage fluctuations across each junction are measured simultanerously. The right inset shows a schematic of the crossed Andreev reflection process.}
\end{figure}

The nonlocal resistance of hybrid SN devices has been intensively investigated~\cite{Beckmann2004,Russo2005,Cadden-Zimansky2006,Kleine2009}, but the observation of CAR is complicated by another nonlocal process called elastic cotunneling (EC), in which electrons in the normal-metal leads tunnel across the superconductor with the help of Cooper pairs, resulting in a positive nonlocal resistance and a negative current-current correlation. Theoretical studies have found that the two nonlocal processes tend to cancel each other exactly in the lowest order approximation in the tunneling limit~\cite{Falci2001}, but experimental results and further theoretical investigation show that exact cancellation does not occur if the normal-metal leads are ferromagnetic~\cite{Beckmann2004}, if the NS interfaces are highly transparent~\cite{Cadden-Zimansky2006,Kalenkov2007b}, or if there is strong Coulomb interaction~\cite{Russo2005,Yeyati2007}. For tunneling junctions the Coulomb interaction may lead to a transition from EC to CAR with increasing voltage bias~\cite{Russo2005,Yeyati2007}. 

The noise cross-correlation measurement is an especially powerful tool to probe the correlation between charge carriers in mesoscopic systems~\cite{Blanter2000}. For example, cross-correlation measurements in the Hanbury Brown and Twiss type experiment have demonstrated the Fermionic nature of electrons~\cite{Oliver1999,Henny1999b}, and have been used for probing orbit entanglement of electrons from independent sources~\cite{Neder2007nature,Neder2007prl}. It was also predicted that current-current correlation can be used to directly probe CAR and EC without the drawbacks of the resistance measurement~\cite{Bignon2004,Melin2008}. However, despite considerable amount of theoretical work, no experimental observation has been reported. Below we describe both nonlocal resistance and noise cross-correlation measurements, and a clear signature of entanglement is observed in each type of measurement.

In Fig.~1 we show a scanning micrograph of  a hybrid device, as well as the schematic of the measurement circuit (see the Methods section for sample fabrication and measurements). Two normal-metal/insulator/superconductor (NIS) junctions are formed between Al and Cu leads, with room temperature resistances in the range of 10--20 k$\Omega$. The junction size is about 0.3 $\mu$m by 0.45 $\mu$m, and the distance between the two junctions is about 0.28 $\mu$m, comparable to the superconducting coherence length $\xi_{S}$ of diffusive Al. Three batches of devices were measured and consistent results were obtained. Here we concentrate on a single device with the most complete data. The advantage of using tunneling junctions instead of junctions with highly transparent interface is the following: First,  the large tunneling resistance ensures there is little leakage current from one junction to the other, i.e., currents from different sides only flow to the superconducting lead (ground), so the bias across the two junctions can be varied independently. Second, for a highly transparent interface, EC is expected to dominate over CAR at all temperatures and at energies below the gap voltage~\cite{Cadden-Zimansky2006,Kalenkov2007b}, while for tunneling junctions CAR can dominate EC at high bias~\cite{Russo2005,Kleine2009,Yeyati2007}. Third, the Coulomb interaction between the electrons in the normal metals may lead to a   ``dynamical'' Coulomb blockade effect (DCB), which prevents both electrons from being injected into the same normal-metal lead and thus facilitates the separation of the two spin-entangled electrons~\cite{Recher2003}. Note that this is different from the DCB in the superconducting lead which favors EC instead of CAR~\cite{Yeyati2007}. Fourth, tunneling junctions can be treated exactly in perturbation theory, which makes the comparison between theoretical predictions and experimental results straightforward. Finally, the fact that the junction resistance is much larger than that of the normal-metal lead ensures a sensitive measurement of current across the junction, as voltage is the quantity actually measured here.

\begin{figure}
\includegraphics[width=6in]{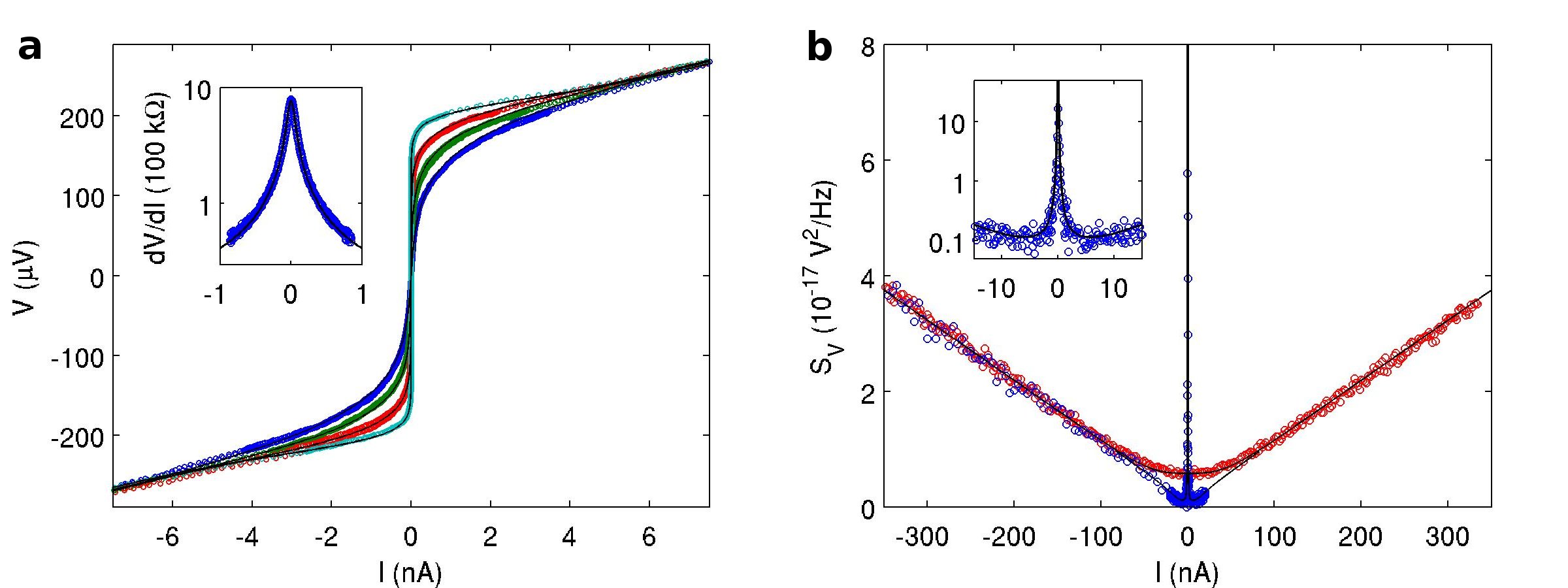}
\caption{ Junction characterization. \textbf{a}, Current voltage characteristics of an SN junction at bath temperatures: 0.1 K (cyan), 0.2 K (red), 0.3 K (green), and 0.4 K (blue). Inset: Bias dependence of the differential resistance measured at 0.4 K (note the log scale of the y axis). The solid lines are fits using the semiconductor model, with slightly elevated electron temperature.  \textbf{b}, Bias dependence of the voltage noise power of an SN junction at 4.2 K (red) and 0.4 K (blue). Inset: Zoomed in regime for the 0.4 K data near zero bias. The solid lines are fits to the shot noise expression (Fano factor equals to 1) with background noise from the measurement setup also taken into account. The diverging behavior near zero bias is due to the large differential resistance as shown in \textbf{a}.  }
\end{figure}

The properties of our NIS junction,  as shown by the current voltage characteristics (CVCs) at different temperatures in Fig.~2a, can be understood by a simple semiconductor model~\cite{Tinkham1996}.  Below the superconducting transition temperature, the number of states in the superconductor available for quasiparticle tunneling decreases exponentially, leading to a  diminishing current at low bias voltage. The ratio between the differential resistance at zero bias and the normal state junction resistance is $(\sqrt{k_{B}T/2\pi \Delta})e^{\Delta/k_{B}T}$ in the limit $k_B T\ll \Delta$. If there is no pin hole in the thin insulating layer, current flowing across the NIS junction consists of well separated events of quantum mechanical tunneling of electrons (no temporal correlation and inelastic scattering). In this case, the power of the current fluctuation is proportional to the magnitude of the current. This so-called shot noise reflects the particle nature of the charge carriers~\cite{Blanter2000}. Here the voltage noise power $S_V$ is measured  and it is related to the current noise power $S_I$ by 
$S_{V}=S_{I}\cdot (\frac{dV}{dI})^{2}$, with 
\begin{equation}
S_{I}=\frac{4(1-F)k_{B}T}{R}+F\cdot 2eI \coth(\frac{eV}{2k_{B}T}),
\end{equation}  
where $dV/dI$ is the differential resistance at bias voltage $V$, and $F$ is the Fano factor, which equals 1 for full shot noise~\cite{Buttiker1992,Blanter2000}. At  $eV \gg k_B T$, the hyperbolic cotangent term approaches 1 and $S_I$ is proportional to $I$, the hallmark of shot noise, as shown in Fig.~2b for our sample. Below the superconducting transition temperature, Andreev reflection may lead to doubled shot noise near zero bias, as has been observed for short diffusive NS junctions with transparent interfaces~\cite{Jehl2000}, for which the differential resistance at zero bias is smaller than the normal resistance. However, for typical NIS junctions like ours, the probability of AR is very small~\cite{Eiles1993,Pothier1994}, thus the subgap current due to AR (the only charge transfer process at zero temperature) is also small. As the shot noise is proportional to the current, it is difficult to verify the doubled shot noise for NIS junctions~\cite{Lefloch2003}. Although CAR can also be inferred from shot noise  (auto-correlation)  measurements~\cite{Bignon2004}, observing CAR by shot noise measurements is even harder since CAR is usually dominated by the AR process, and its amplitude decays exponentially with the distance between the two normal metals~\cite{Bignon2004,Cadden-Zimansky2006}. Another practical issue for auto-correlation measurements is that the extrinsic current noise from the measurement setup (back action) can be much larger than the intrinsic shot noise. 

\begin{figure}
\includegraphics[width=6in]{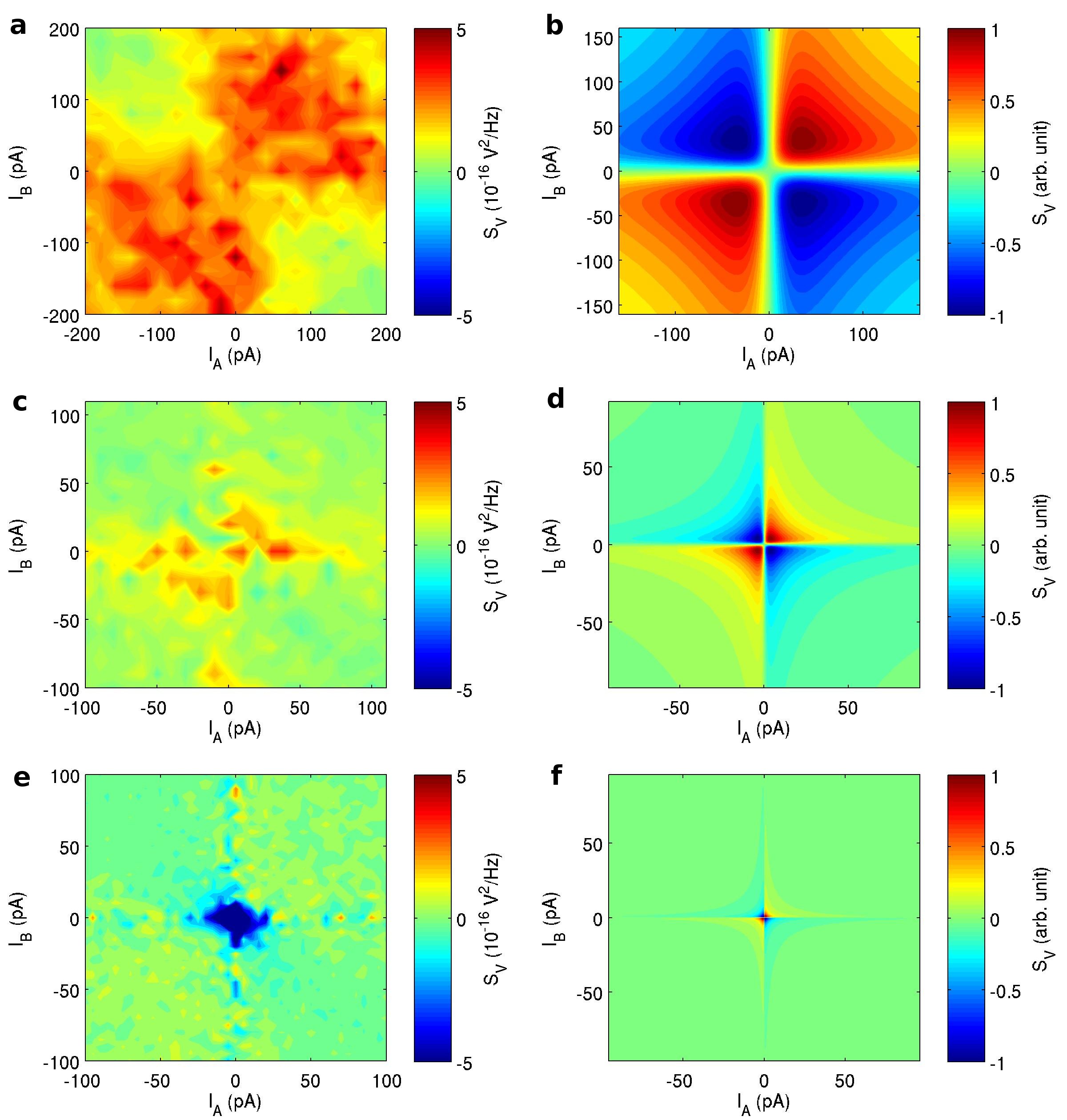}
\caption{ Bias current dependence of voltage noise power by the cross-correlation measurement. The left column shows the measured voltage noise power at three different bath temperatures: 0.4 K (\textbf{a}), 0.3 K (\textbf{c}), 0.25 K (\textbf{e}). The right column shows the respective  theoretical prediction at each temperature, with the assumption $A^{CAR}=A^{EC}$ in Eq.(2).}
\end{figure}

Fortunately, with the nonlocal configuration shown in Fig.~1, the cross-correlation between $V_A$ and $V_B$ is only determined by nonlocal processes, independent of shot noise of both junctions. In addition, it should not be affected by any extrinsic current noise, as there is no circuit segment shared between the two channels except a short piece of superconducting Al wire, along which no voltage drop is expected to occur. Moreover,  the large differential resistance near zero bias (see Fig.~2) leads to a large nonlocal voltage-voltage correlation signal which greatly helps the observation of the small nonlocal current-current correlation. However, the large differential resistance here also limits the bandwidth used for the cross-correlation measurement due to the capacitance of the electrical wiring (see the Methods section), which prevents us from making a reliable measurement below about 0.2 K. 

Figure~3 shows the voltage noise power measured by cross-correlation, $S_V$, for our device at three different temperatures: 0.4 K, 0.3 K, and 0.25 K. For SN tunneling junctions in the abscence of the Coulomb interaction in the leads, it was shown that when $e|V|$, $k_B T \ll \Delta$, the current-current correlation between junction $A$ and junction $B$ is~\cite{Bignon2004}
\begin{equation}
  S_{AB}
  =S^{CAR}-S^{EC}=
  2e G_Q \,
  \left[
  (  V_{A} +  V_{B} )
  \coth \left( \frac{ e V_{A} + e V_{B}} {2 k_{B}T}\right)
  A^{CAR}
  - (  V_{A} -  V_{B} )
  \coth \left( \frac{ e V_{A} - e V_{B}}{2 k_{B}T}\right)
  A^{EC}
  \right]
\end{equation}
where $G_{Q}=2e^{2}/h$, $A^{CAR}$($A^{EC}$) is the amplitude of CAR(EC), determined by the distance between the two SN junctions and the properties of the interfaces. For simple tunneling junctions, $A^{CAR}=A^{EC}$ is predicted~\cite{Falci2001}. Since the precise value of $A^{CAR}$ is not known, a normalized $S_{AB}$ is plotted in Fig.~3 to compare with experimental results.  It is easy to see that by setting $V_A =V_B $, the EC term in the bracket of Eq.(2) reduces to a bias independent term $2k_B T A^{EC}$, and the CAR term is maximized. Since current sources instead of voltage sources were used in the experiment, current voltage characteristics from the semiconductor model (see Fig.~2) were used to convert $S_{I}(V_{A},V_{B})$ to $S_{V}(I_{A},I_{B})$ for comparison.

At all three temperatures, when the bias of the two junctions is of the same polarity, there is a clearly positive correlation, a signature of entanglement, consistent with the theoretical prediction. When the bias of the two junctions is of opposite polarity, Eq.~(2) predicts a negative $S_V$.  In the experiment, $S_V$ changes from slightly positive to slightly negative as temperature decreases, again consistent with predictions, but also suggesting a slowly increasing amplitude of the EC process, which means  the assumption $A^{CAR}=A^{EC}$ at finite temperature is oversimplified.
At $V_A ,V_B \rightarrow 0$, while $S_{AB}$ is predicted to be $2k_B T (A^{CAR}-A^{EC}$), the experimental data show a positive correlation at 0.4 K that evolves to a sharp negative dip at 0.25 K. Such a sharp negative correlation again indicates that the EC process strongly dominates over the CAR process at lower temperature, which is not consistent with theoretical predications. 

To understand this discrepancy, Coulomb interaction in the leads need to be considered~\cite{Recher2003,Yeyati2007,Melin2008}. We note that the charging effect of solitary junctions has a strong temperature and bias dependence~\cite{Kauppinen1996prl}. As the Coulomb energy associated with charging the superconducting lead and charging the normal-metal leads changes, $A^{CAR}$ and $A^{EC}$ also change and this may lead to a nonmonotonic temperature and bias dependence of $S_{AB}$. Note that here the junction capacitance is much smaller than that of the planar NISIN devices studied before~\cite{Russo2005}. Since the charging energy is $e^{2}/2C$, where $C$ is the capacitance of the tunneling junction, a much stronger DCB effect is expected. The effect of DCB on noise correlation was recently  considered with a renormalization group approach~\cite{Melin2008}, where the presence of Coulomb interaction in the normal-metal leads is treated as a reduction of interface transparency, resulting in a change of the sign and magnitude of the correlation.  Although this approach is more general than the analytical result in Eq.~(2), it does not seem to be applicable as the Coulomb interaction in the superconducting lead was not taken into account in the theory. Further experiments with a modified local impedance near the normal-metal leads that enhances the DCB effect may facilitate comparison with theory~\cite{Recher2003,Melin2008}.  We note also that in most theoretical work only pair tunneling (AR, CAR, and higher order processes) is considered in the zero temperature limit, while in the experiment, quasiparticle tunneling dominates at finite temperatures since a reasonable differential resistance is required for measurement.

\begin{figure}
\includegraphics[width=6in]{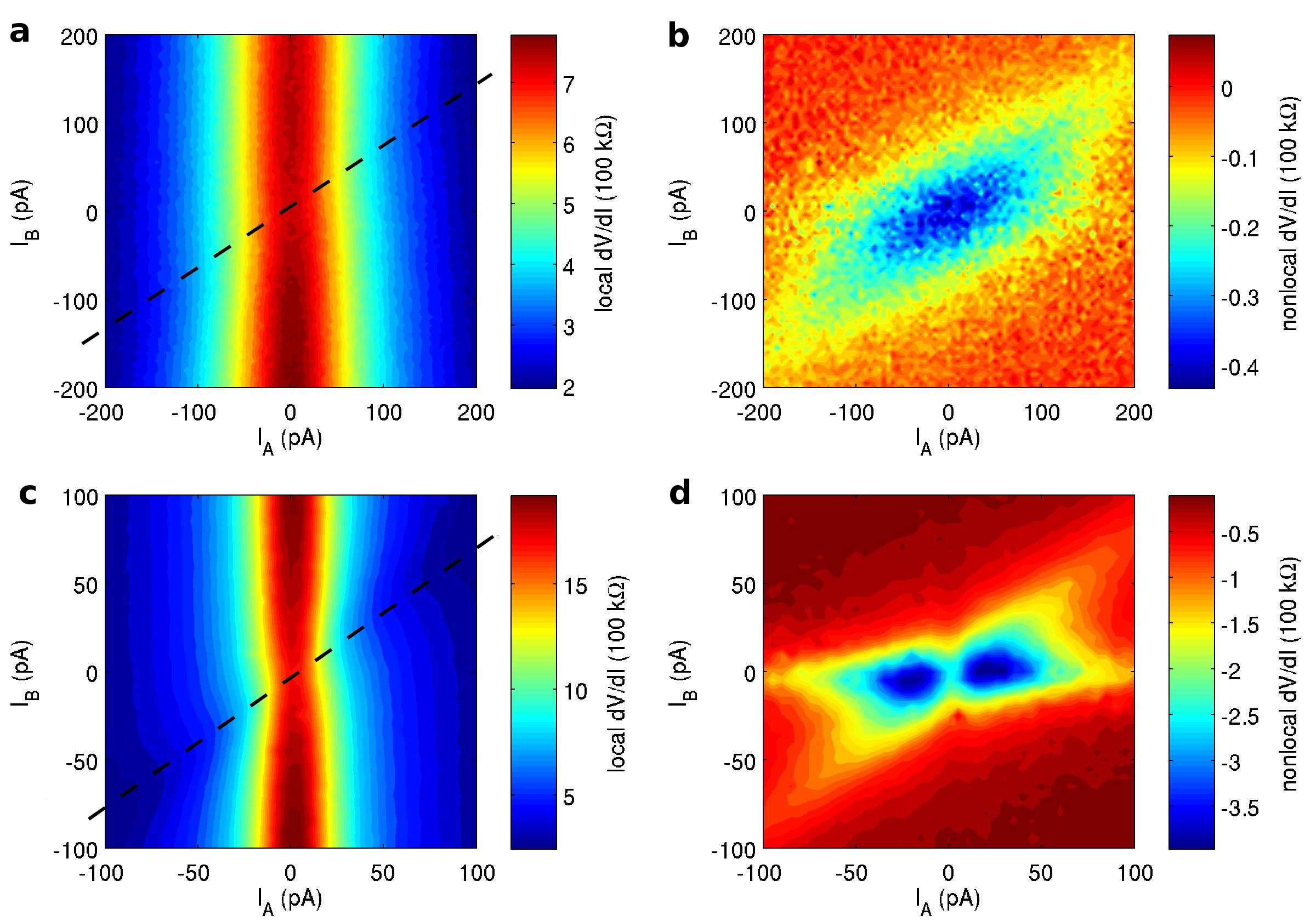}
\caption{ Bias current dependence of local and nonlocal differential resistance. The left column shows the local differential resistance at bath temperatures 0.4 K (\textbf{a}) and 0.3 K (\textbf{c}), obtained by adding a small ac current modulation to $I_A$ and measuring the ac voltage modulation of $V_A$. The nonlocal resistance is obtained at the same time by measuring the voltage modulation of $V_B$, as shown in the right column at 0.4 K (\textbf{b}) and 0.3 K (\textbf{d}). The black dashed lines indicates a  strong reduction of the local differential resistance, approximately where the dc bias voltage is the same for the two junctions (the normal state resistances of the two junctions are 15 and 19 k$\Omega$ respectively).}
\end{figure}

The local and nonlocal differential resistance corroborate the observation from the noise cross-correlation measurements: They also demonstrate a clear signature of entanglement, as shown in Fig.~4. As noted before, since the tunneling junction resistance is large, there is little current redistribution (flow from one normal-metal lead to the other via the superconducting lead), which enables us to do a 2D scan of both local and nonlocal differential resistance, in contrast to previous 1D nonlocal resistance measurements~\cite{Beckmann2004,Russo2005,Cadden-Zimansky2006,Kleine2009}. Surprisingly, even for the local resistance, there is clearly a reduction of resistance when $V^{dc}_{A} \approx V^{dc}_{B}$, as indicated by the dashed line in Fig.~4. This strong  reduction is not expected as the differential resistance is determined by quasiparticle tunneling, and even if there is a small probability of pair tunneling, direct AR is expected to dominate CAR and EC~\cite{Falci2001,Bignon2004}. However, this might suggest that DCB indeed enhances the  splitting of a Cooper pair into separate normal-metal leads, especially when the electrons passing the two SN interfaces are of the same energy, in agreement with predictions~\cite{Recher2003}.

Figure 4b shows that at 0.4 K, the nonlocal resistance is negative in the subgap regime for both junctions, a clear signature of spin entanglement. As the temperature decreases to 0.3 K, a peak evolves near zero bias (at the peak, the nonlocal resistance even becomes positive at 0.25 K, data not included here). Remarkably, if we take the 1D cut at $I_{B} = 0$, the bias dependence of the nonlocal resistance is close to that previously reported where only one junction was biased~\cite{Russo2005,Kleine2009}. This confirms that the nonlocal resistance measured in the current setup has the same origin; though a more completed picture can be obtained here.

In summary, we have shown experimental evidence of spin entanglement revealed by noise cross-correlation measurements between two spatially separated normal-metal leads. We find a clear positive correlation when the bias of the two junctions is of the same polarity, indicating the nonlocal Andreev reflection and spin entanglement of electrons is maximized, consistent with theoretical predictions. At lower temperature and near zero bias, we find a strongly negative correlation, possibly due to a dynamical Coulomb blockade effect in the superconducting lead. We also find that both local and nonlocal differential resistance measurements can be used to demonstrate that the entanglement is maximized with the same bias voltage across the two SN interfaces. These findings may lead to better understanding and  control of  quantum entanglement devices made of hybrid SN structures. Further experiments with spin selective normal-metal leads and Bell inequality type measurements~\cite{Chtchelkatchev2002,Melin2008},  will provide a more conclusive  check of spin entanglement.

\textbf{Methods}

The devices were fabricated using standard two angle electron-beam lithography and e-gun evaporation. A polymethyl methacrylate/polydimethyl glutarimide (PMMA/PMGI) bilayer was spin-coated on Si substrates with 300 nm SiO$_{2}$ insulating layer for patterning the devices. For the particular device reported here, a 27 nm 99.999\% pure Al film was deposited first in an e-gun evaporator with a base pressure of 2.2$\times 10^{-7}$ Torr at a 40 degree angle and at a rate of 0.1 nm/sec. After deposition, 0.2 Torr pure oxygen gas was allowed into the chamber for about 5 minutes to create a thin layer of oxide. Then a 50 nm of Cu film was deposited at a normal angle. Devices from three batches were measured in an Oxford KelvinOx 100 or an Oxford KelvinOx 300 dilution refrigerator. Results obtained on these devices were consistent with each other.

The current sources shown in Fig.~1 were realized by putting a large ballast resistor (1 G$\Omega$) in series with a filtered voltage source. The ac voltage signals from the junctions were amplified using home-made battery-powered low noise amplifiers. The voltage signals after amplification were sent to data acquisition cards and were analyzed by a computer after digitization. To reduce noise coupled to devices at low temperature, $\pi$-filters on top of the cryostat were used. The measurement signals from the cryostat were amplified inside a $\mu$-metal enclosure close to the cryostat. The bandwidth of measurement was limited by the roll-off effect due to the capacitance of the $\pi$-filter (about a few nF). 

The auto-correlation data shown in Fig.~2b at 4.2 K were measured in the frequency range from 500 to 1000 Hz, while at 0.4 K different  frequency ranges, from 200 to 600 Hz and from 120 to 130 Hz, were used and similar results were obtained. The voltage noise power $S_{V}$ is fitted to,
\begin{equation}
 S_{V}=2eI \coth(\frac{eV}{2k_{B}T}) (\frac{dV}{dI})^{2}+2(I_{n}\frac{dV}{dI})^{2}+V_{n}^{2},
\end{equation} 
where  $I_{n}$ is the extrinsic current noise, and $V_{n}$ is the extrinsic correlated voltage noise. A good fit with data was found by using  $I_{n}=$30 fA/$\sqrt{Hz}$, and $V_{n}=$0.7 nV/$\sqrt{Hz}$. These values are consistent with standard JFET amplifiers. Hence, in the sub-gap regime, the extrinsic current noise dominates the intrinsic shot noise.

The cross-correlation data shown in Fig.~3 were obtained in the frequency range from 2 to 6 Hz to avoid the roll-off effect and the side lobe of the 60 Hz line frequency peak. The efficiency of the cross-correlation measurement at this frequency range was compromised by the 1/f noise of amplifiers. Nevertheless, due to the large differential resistance the bias dependence of $S_{AB}$ is clearly visible.

\end{document}